\documentstyle[prl,aps,preprint]{revtex}
\begin{document}
\draft
\title{Generation of spin-polarized currents
in Zeeman-split Tomonaga-Luttinger models}
\author{Takashi Kimura, Kazuhiko Kuroki, and Hideo Aoki}
\address{Department of Physics, University of Tokyo,
Hongo, Tokyo 113, Japan}
\date{\today}
\maketitle
\begin{abstract}
In a magnetic field an interacting electron gas in one dimension
may be described as a Tomonaga-Luttinger model
comprising two components with different
Fermi velocities due to the Zeeman splitting.
This destroys the spin-charge separation, and
even the quantities such as the density-density correlation
involve spin and charge critical exponents ($K$).
Specifically, the ratio of the up-spin and down-spin
conductivities in a dirty system
{\it diverges} at low temperatures
like an inverse power of the temperature,
$T^{-(K_{\uparrow}-K_{\downarrow})}$,
resulting in a spin-polarized
current.
In finite, clean systems the conductance
becomes different for up- and down-spins as another manifestation of
the electron-electron interaction.

\end{abstract}

Recent studies of quantum transport in mesoscopic
systems have brought to light many unusual features
unexpected for classical systems \cite{meso}.
This is heightened by recent advances in fabricating
nanostructure quantum wires and quasi-one-dimensional
(1D) crystal structures.

In 1D systems,
 the interactions between the electrons is so crucial
due to the strong constraint in the phase space that
the system becomes universally what is called
the Tomonaga-Luttinger (TL) liquid
as far as low-lying excitations are concerned
no matter how small the interaction may be\cite{Lutt}.
A most striking feature of this 1D model
is the spin-charge separation.
The transport properties\cite{1D}
are also dominated by the spin-charge
separation in the following sense.
The low-temperature
 conductivity of the dirty TL liquid
as studied by Luther and Peschel\cite{Luther}
 exhibits a power law, $\sigma(T)\sim T^{2-K_{\rho}-K_{\sigma}}$.
The power law comes from the degraded
Fermi singularity in the TL model,
while the critical exponents (which are
functions of the interaction) enter as a sum of $K_{\rho}$ for
the charge phase of the system and
$K_{\sigma}$ (which is actually fixed at
$1$ for spin-independent interactions) for the spin phase.
 A recent experiment
\cite{Tarucha} for high-quality quantum wires seems
 to support this result.
 For clean systems
 Kane and Fisher \cite{KF}
 and Furusaki and Nagaosa \cite{FN} found that
 the conductance quantization in finite systems
in the noninteracting case
 becomes proportional to the exponent,
 $G=(e^2/\pi)K_{\rho}$
 (where $\hbar=k_B=1$ is assumed hereafter).

 Now we can raise an intriguing question: what happens
if we degrade the spin-rotational (SU(2)) symmetry?
Such a situation is realized
by applying a magnetic field, which makes the
 Fermi velocities spin dependent
due to the Zeeman splitting.
In this paper,
we show that the spin-charge separation will be then
destroyed, causing
even the quantities such as the density-density correlation
involve spins.
Thus the spin may manifest itself
in the transport, leading possibly to spin-polarized currents,
which is shown to be the case.

The generation of spin-polarized currents
has been of a long-standing
 interest for academic\cite{Edel} as well as
practical points of view, where
typical applications include
spin-polarized STM\cite{wiesen} and the
Mott-detector\cite{koike}.
Fasol and Sakaki \cite{Fasol}
have suggested that in the spin-orbit split
bands of GaAs quantum wires
the curvature in the band dispersion (as opposed to the linearized
dispersion in the Tomonaga-Luttinger model)
will make the
relaxation time due to the electron-electron interaction
spin-dependent and consequently make
the outgoing current spin-polarized.
The mechanism proposed in this Letter is by
contrast a purely electron-correlation effect,
where the ratio, $\sigma_{\uparrow}/\sigma_{\downarrow}$, diverges
toward $T=0$.

We start from a clean, two-band Luttinger model,
which is similar to the one employed
in a study for the Fermi-edge singularity
in 1D\cite{Ogawa}.  The Hamiltonian is given by
\begin{eqnarray}
H_{\rm clean}&=&H_0+H_{\rm int} ,
\end{eqnarray}
where the non-interacting Hamiltonian $H_{0}$
is written as
\begin{eqnarray}
H_0&=&\sum_{k,s,i}v_{Fs}[(-)^{i+1}k-k_{Fs}]
     c^{\dagger}_{iks}c_{iks}\nonumber\\
&=&\frac{2\pi}{L}\sum_{q>0}\sum_i\{ v_0[
       \rho_{i}(q)\rho_{i}(-q)+\sigma_i(q)\sigma_i(-q)]\nonumber\\
    &&+\Delta v[\rho_i(q)\sigma_i(-q)+\rho_i(-q)\sigma_i(q)]\} .
\end{eqnarray}
Here $v_{F\uparrow}$ ($v_{F\downarrow}$)
is the Fermi velocity of the up (down)
spin subband in a magnetic field $B$ with an average
$v_0 \equiv (v_{F\uparrow}+v_{F\downarrow})/2$ and the difference
$\Delta v \equiv (v_{F\uparrow}-v_{F\downarrow})/2$.
Similarly,
$k_{F\uparrow}$ ($k_{F\downarrow}$) is the Fermi momentum of the
 up (down) spin, which equal to $k_F$ in the absence of $B$.
$c_{iks}^{\dagger}$ creates
a right-going ($i=1$) or left-going ($i=2$) electron
 with momentum $k$ and spin $s$, while
$\rho_i(q)$ [$\sigma_i(q)$] are the usual charge [spin]
density operators, and $L$ is the system size.

The second line in the above equation indeed indicates that
the charge and spin
are no longer decoupled for $\Delta v \neq 0$,
in a sharp contrast to the usual
Tomonaga-Luttinger model.

The spin-charge separated part can be cast into the usual
phase Hamiltonian where $v_0$
corresponds to the Fermi velocity $v_F$.
Thus we can
introduce the usual phase field
$\theta_+(x)$ ($\phi_+(x)$) and
the dual field $\theta_-(x)$ ($\phi_-(x)$)
 corresponds to the charge (spin) degree of freedom to have
\begin{eqnarray}
H_{\rm clean}&=&\ \ \frac{v_{\rho}}{4\pi}\int\ dx\Big\{
        \frac{1}{K_{\rho}}[\partial_x\theta_{+}(x)]^2+
        K_{\rho}[\partial_x\theta_{-}(x)]^2\Big\}\nonumber\\
      &&+\frac{v_{\sigma}}{4\pi}\int\ dx\Big\{
       \frac{1}{K_{\sigma}}[\partial_x\phi_{+}(x)]^2+
        K_{\sigma}[\partial_x\phi_{-}(x)]^2\Big\}\nonumber\\
      &&+\frac{\Delta v}{2\pi}\int\ dx
        \Big\{[\partial_x\theta_{+}(x)][\partial_x\phi_{+}(x)]
        +[\partial_x\theta_{-}(x)][\partial_x\phi_{-}(x)]\Big\},
\end{eqnarray}
where $K_{\rho}$ ($K_{\sigma}$) is the critical exponent
of the charge (spin) phase.
In the following we assume that
the coupling constants between electrons do not have
 insignificant magnetic field dependences, so that we have
$K_{\sigma}=1$, $v_{\sigma}=v_0$,
$K_{\rho}=1/\sqrt{1+4g}$, and
$v_{\rho}=v_0/K_{\rho}=v_0\sqrt{1+4g}$,
where $g(\sim U/2\pi v_0$ for the Hubbard model)
is the dimensionless,
forward-scattering coupling constant.
Here we have neglected the backward scattering and Umklapp scattering
 processes, since
 they have large momentum transfers.

We can diagonalize $H_{\rm clean}$, as
is done for the electron-hole system in a
two-channel Tomonaga-Luttinger study of the excitonic
phase by Nagaosa and Ogawa\cite{nagaosa}, via a
linear transformation to two new phases,

\begin{eqnarray}
\left( \begin{array}{c}
\theta_+(x) \\ \phi_+(x)
\end{array} \right)
=
\left( \begin{array}{cc}
{\rm cos}\alpha & -\frac{1}{y}{\rm sin}\alpha\\
y{\rm sin}\alpha & {\rm cos}\alpha
\end{array} \right)
\left( \begin{array}{c}
\tilde{\theta}_+(x) \\ \tilde{\phi}_+(x)
\end{array} \right)
\end{eqnarray}
where $\alpha$ is the
`rotation angle in the spin-charge space'
($\propto \Delta v$ for small $\Delta v$) with
${\rm tan}2\alpha=2(\Delta v/v_0)
\sqrt{2(K_{\rho}^{-2}+1)}/(K_{\rho}^{-2}-1)$ and
$y^2=\frac{1}{2}(K_{\rho}^{-2}+1)$.
These new phases also have gapless, linear dispersions,
in which the new velocities are given by
\begin{eqnarray}
\tilde{v}_{\rho,\sigma}^2&=&\Delta v^2+\frac{1}{2}v_0^2\Big[
       K_{\rho}^{-2}+1\pm (K_{\rho}^{-2}-1)\sqrt{
       1+{\rm tan}^{2}2\alpha}\Big],\\
\end{eqnarray}
where $+(-)$ sign corresponds
to $\tilde{v}_{\rho} (\tilde{v}_{\sigma})$.

Now we can turn to the calculation
of the conductivity in a dirty system.
The total Hamiltonian is then $H = H_{\rm clean}+H_{\rm imp}$, where
the impurity scattering part $H_{\rm imp}$ is
given by
\begin{eqnarray}
    H_{\rm imp}\,&=&\,\sum_{s}\sum_l \int \!\! dx \
              N_{s}(x) \,u(x-x_l)\,,
\end{eqnarray}
where $u(x-x_l)$ is the
impurity potential situated at $x_l$ and $N_{s}(x)$
is the density operator of spin $s$ electrons, whose
phase representation is
$N_s=\frac{1}{2\pi}{\partial_x(\theta_++s\phi)}+\frac{1}{\pi \Lambda}
{\rm cos}[2k_{Fs}x+\theta_++s\phi]$
with $\Lambda$ being a short-range cutoff.
The conductivity of
$s$-spin subband, $\sigma_s$, is given by $\sigma_s=n_e
e^2\tau_s/2m_s^*$, where $\tau_s$ is the relaxation time
of the spin $s$ subband,
$n_i$ ($n_e$) is the density of impurities (electrons),
 and $m_s^* \propto v_s^{-1}$ is
the effective mass of the spin $s$ subband.
In 1D we have $n_e=2k_F/\pi$, but we have neglected the trivial
magnetic-field dependence of
$k_{F\uparrow}$ and $k_{F\downarrow}$
to single out the effect of differentiated $v_{F\uparrow,\downarrow}$.

We can then calculate $\tau_s$ following
G{\"o}tze and W{\"o}lfle in the Mori formalism
for the conductivity\cite{gotze}\cite{TKAE},
in the second order in $H_{\rm imp}$ as

\begin{eqnarray}
\frac{1}{\tau_s}&\approx&4\pi v_{Fs}n_iu^2(2k_F)\sum_q
                 \lim_{\omega\rightarrow
                   0}\frac{{\rm Im}\Pi_s(2k_{Fs}+q,\omega)}{\omega},
\end{eqnarray}
where $u(q)$ is the Fourier transformation of $u(x)$, and $\Pi_s$ is
the density-density correlation function.

In terms of the density operator
$\rho_s(x)$ for spin $s$ we have
\begin{eqnarray}
\lim_{\omega\rightarrow 0}\sum_q\frac{{\rm Im}\Pi_s(2k_{Fs}+q)}{\omega}
    =\frac{1}{2T}\sum_{s^{\prime}}\int_{-\infty}^{\infty}\ dt\
     \langle\rho_s(0,t)\rho_{s^{\prime}}(0,0)\rangle.
\end{eqnarray}
In the summation over the spin $s'$
we can readily
show that the cross term,
$\langle \rho_{\uparrow}(0,t) \rho_{\downarrow}(0,0) \rangle$, vanishes.
Then
the conductivity becomes
a sum of the two spin components, each of which has
a simple power-law temperature
dependence as in the usual Luttinger theory,
\begin{eqnarray}
\sigma_{s}(T)=\sigma_0\Big(\frac{v_{Fs}}{v_0}\Big)^2
         \Big(\frac{T}{\omega_F}\Big)^{2-K_s},
\end{eqnarray}
where $\sigma_0\equiv \sigma(T=\omega_F)$ (whose dependence on
$k_{Fs}$ is again ignored here)
and $\omega_F\sim \epsilon_F$ is the high-energy
cutoff.

Here the spin-dependent exponent $K_s$ is given
by
\begin{eqnarray}
K_{s}
   =({\rm cos} \alpha+sy{\rm sin} \alpha)^2 \tilde{K}_{\rho}
    +({\rm cos} \alpha - \frac{s}{y}{\rm sin} \alpha)^2 \tilde{K}_{\sigma},
\end{eqnarray}
where $\tilde{K}_{\rho(\sigma)}$ are the
 critical exponent of the phase $\tilde{\theta}$ ($\tilde{\phi}$)
given by
\begin{eqnarray}
\tilde{K}_{\rho(\sigma)}^2&=&y^{\mp 2}\Big[K_{\rho}^{-2}+3
         \pm (K_{\rho}^{-2}-1)
\sqrt{1+{\rm tan}^{2}2\alpha}\Big]\nonumber\\
                          &&\times
\Big[3K_{\rho}^{-2}+1\pm (K_{\rho}^{-2}-1)
          \sqrt{1+{\rm tan}^{2}2\alpha}\Big]^{-1},
\end{eqnarray}
where the upper (lower) sign corresponds to $\tilde{K}_{\rho}$
($\tilde{K}_{\sigma}$).

The above equations (10)$\sim$(12)
are the key result of this paper:
the electron-electron interaction does indeed make the conductivity
dependent on the spin, where
the power-law dependence in $T$ is retained so that
the spin dependence becomes more enhanced at lower
temperatures.
The ratio
$\sigma_{\uparrow}/\sigma_{\downarrow}
\propto T^{-(K_{\uparrow}-K_{\downarrow})}$
actually diverges toward $T\rightarrow 0$,
resulting in a spin-polarized current.
This divergence is purely an effect of the electron correlation.

The temperature dependence of $
\sigma_{\uparrow}/\sigma_{\downarrow}$ numerically
calculated in the presence of a fixed electron-electron interaction
$g$ and the several ratios of $v_{F\uparrow}$ to
$v_{F\downarrow}$ is displayed in Fig.1.
Figure 2 shows a drastic dependence of
$\sigma_{\uparrow}$ ($\sigma_{\downarrow}$)
 on the ratio $v_{F\uparrow}/v_{F\downarrow}$
at a fixed temperature with a fixed $g$.

The result shows that the more conductive channel is the spin having
a smaller $v_F$,
since $K_{\uparrow}>K_{\downarrow}$ for
$\Delta v \propto (v_{F\uparrow}-v_{F\downarrow}) >0$.
Physically,
the effective electron-electron interaction is
smaller (larger) in the lighter (heavier)
spin subband, since it is the ratio of the
electron-electron coupling constant to the
kinetic energy ($\propto v_F$)
that matters.
Thus the result is roughly consistent with the observation in a
single TL liquid that the electron-electron repulsive
 interaction suppresses the conductivity\cite{Luther}.

If we look more closely for an intuitive
interpretation of the present result,
one could regard the single TL model of spin 1/2 electrons as
a double-chain system of spinless electrons.
The present situation
could then be regarded as a generalization of our previous
model\cite{TKAE}, where we have considered two equivalent chains
having intrachain and interchain interactions
 in the absence of interchain tunneling.
When the two `chains' are made inequivalent by the differentiated
$v_F$, this modifies both the `intra-chain'
(parallel-spin) dimensionless coupling constant,
$g/v_F$, and
`inter-chain' (antiparallel-spin) $g/v_F$.
Note that, if the $g$-parameters derive from an SU(2)
symmetric interaction (such
as the Hubbard $U$) the $g$'s for intra- and inter-chain interactions are
the same.  It is then a highly nontrivial question what
conductivities will come out.
The present result indicates that the chain that has a smaller
$g/v_F$ does indeed remain more conductive, so that
the effect of parallel-spin interaction eventually prevails.
This is consistent with the double-chain result that
the intra-chain repulsion suppresses the conductivity, while
the effect of inter-chain interactions,
which incidentally enhances the conductivity, is only of the
second order.

Next we consider the {\it conductance} of finite, clean systems.
We can calculate the conductance, $G_s$, of spin $s$ subband
from the current-current correlation function as \cite{KF}
\begin{eqnarray}
G_s=\lim_{\omega \rightarrow 0}\sum_{s^{\prime}}\frac{1}{\omega L}
\int d\tau\int dx\ e^{i\omega \tau}
\langle T_{\tau}J_s(\tau)J_{s^{\prime}}(0)\rangle,
\end{eqnarray}
where $J_{s} = \partial_{\tau}(\theta+s\phi)/2$
is the spin $s$ current, and $\tau$ is the imaginary
time. Then we end up with
\begin{eqnarray}
G_{s}=\frac{e^2}{2\pi}\Big[
  ({\rm cos}^2\alpha + \frac{ys}{2}{\rm sin} 2\alpha)\tilde{K}_{\rho}
 + (\frac{1}{y^2}{\rm sin}^2\alpha - \frac{s}{2y}{\rm sin} 2\alpha)
 \tilde{K}_{\sigma}\Big].
\end{eqnarray}
Thus the conductance too depends on the spin in
contrast to the ordinary case with
$G_{\uparrow}=G_{\downarrow}=(e^2/2\pi)K_{\rho}$.
We may emphasize that the Fermi velocity of
the system does not appear in the Landauer formula \cite{land},
so that
the full expression of the spin-dependent
conductance is generated by the electron-electron
interaction (while in the conductivity the power of the
temperature is).
Figure 3 numerically depicts the way
in which $G_{\uparrow}$ ($G_{\downarrow}$)
 increase (decrease) with the ratio
$v_{F\uparrow}/v_{F\downarrow}$ with a fixed $g$.

Finally let us make a comment on the Anderson localization.
 In Ref.\cite{GS1} Giamarchi and Schulz have shown
in the absence of
magnetic fields that the temperature at which
 the Anderson localization occurs in 1D is lower for larger $K_{\rho}$.
 Combining this result with our present result in the presence of
 magnetic field, we can envisage a
dramatic situation in which the electrons
with one spin are Anderson-localized
 while the others are conductive in a certain temperature
 region.

We believe these novel many-body effects can be experimentally measured
in quantum wires by taking appropriate fillings
of the up- and down-spin subbands in
 a given magnetic field.
 Unfortunately, in the case of the usual electron-doped GaAs quantum
 wires the g-factor ($\sim-0.4$ in the bulk) is too small to attain
 the sufficient
 Zeeman splitting. However, if we can prepare e.g. InSb quantum wires
 (whose g-factor is as large as $\sim-50$
in the bulk\cite{LB}), the strength of the
 Zeeman splitting in a typical magnetic field of 1T
 amounts to g$\mu_BH\sim$ 3.0meV.
 In such cases
a significant deviation of $v_{\uparrow}/v_{\downarrow}$
 from unity may be expected.


We are much indebted to Professor Gerhard Fasol
for illuminating
 discussions, and to Professor Tetsuo Ogawa for sending us preprints
 prior to publication.

\begin{figure}
\caption{
The result for the temperature dependence of
 the ratio $\sigma_{\uparrow}/\sigma_{\downarrow}$ with
 a fixed interaction $g=0.5$, and with several ratios of
 $v_{F\uparrow}$ to $v_{F\downarrow}$.
}
\label{ratio}
\end{figure}
\begin{figure}
\caption{
The result for the dependence on $v_{F\uparrow}/v_{F\downarrow}$
(keeping $v_{F\uparrow}+v_{F\downarrow}$=constant) of
the conductivity $\sigma_{\uparrow}/\sigma_0$ and
$\sigma_{\downarrow}/\sigma_0$ with a fixed temperature
$T=10^{-3}\omega_F\sim100{\rm mK}$ and with a fixed interaction
 $g=0.5$.
}
\label{sigma}
\end{figure}
\begin{figure}
\caption{
The result for the dependence on $v_{F\uparrow}/v_{F\downarrow}$
(keeping $v_{F\downarrow}+v_{F\uparrow}$ constant) of
the conductance $G_{\uparrow}$ and $G_{\downarrow}$ normalized by
 $e^2/\pi$ with a fixed interaction $g=0.5$.
}
\label{G}
\end{figure}
\end{document}